\documentclass[a4paper,10pt]{article}
\usepackage{graphicx}
\addtocounter{page}{-1}
\begin{document}
\pagestyle{empty}
\parindent 8mm
\pagestyle{empty}

\vspace*{2cm}
\title{Quantization of the stag hunt game and the Nash equilibrilum}
\author{Norihito Toyota}
\date{Hokkaido Information University,
59-2 Nishinopporo Ebetsu City \\
E-mail:toyota@do-johodai.ac.jp
}


\maketitle
\baselineskip 5mm
\pagestyle{empty}
\begin{abstract}
In this paper I quantize the stag hunt game in the framework proposed by Marinatto and Weber which, is introduced to quntize the Battle of the Sexes game and gives a general quntization scheme of various game theories.
Then I  discuss the Nash equibilium solution in the cases of which starting strategies are taken in both non entangled state and entangled state and uncover the structure of Nash Equilibrium solutions and compare the case of the Battle of the Sexes game. Since the game has 4 parameters in the payoff matrix  has rather rich structure than the Battle of the Sexes game with 3-parameters in the payoff matrix, the relations of the magnitude of these payoff values in Nash Equilibriums are much involuved. This structure is uncovered completly and it is found that the best strategy which give the maximal sum of the payoffs of both players strongly depends on the initial quntum state. As the bonus of the formulation the stag hunt game with four parameters we can discuss various types of symmetric games played by two players by using the latter formulation, i.e. Chicken game. As result some common properties are found between them and the stag hunt game. Lastly a little remark is made on Prisoner's Dillemma.
\end{abstract}
\bf{key words:} 
\it{quantization of game theory, stag hunt game, Nash equilibrium, quntum infor


\rm
\normalsize
\baselineskip 6mm
\pagestyle{empty}
\thispagestyle{empty}

\renewcommand{\thefootnote}{\fnsymbol{footnote}} 
\rm
\normalsize
\baselineskip 6mm
\section{Introduction} 
\renewcommand{\thefootnote}{\fnsymbol{footnote}} 
Recent speed of the study of quntum computation and quantum information processing  is much remarkable\cite{nie} and a few years ago the quantization of game theories have started in context of quntum information\cite{mey,eis,mar}. Two quantization schemes for game theory have been proposed since. One is the scheme by  Eisert et al.\cite{eis} and another by Marinatto et al.{mar}. Many researchers, however, have studied quantization of game thyeory in the frame work of the first formulationis by this time. Games investigated are mainly Prisoner's Dilemma and moreover the Battle of Sexes game, Chicken game and so on. The battle of sexes games which is a symmetric game with $three\; paremeters$ in the payoff bi-matrix are only studied by using the latter formalism\cite{mar}. 

In this article we discuss various types of symmetric games played by two players by using the latter formulation. Especially the stag hunt game is studied in details which has $four \; parameters$ in the synnetric payyoff bi-matrix like many famous games and so we can discuss a general games, i.e. Chicken game,  with four parameters in the same context of the stag hunt game\cite{toyota1}. The special properties different from these famous games with four parameters is that the payoffs of both players in the stag hunt game are same value at all Nash Equilibrium. Some charastic aspects are also discussed for the stag hunt game. 

Our plan in this article is as follows. In section 2, the definition of the stag hunt game are given and the classical Nash Equilibrium are discussed.  We review the latter formalism of quntization for game theory following the paper written by Marinatto et al.\cite{marin} in the section 3.  In the section 4 the stag hunt game is quantized according to the latter formalism\cite{mar} and quntum Nash Equilibrium are given together with the payoff of both playses. The relations of the magnitude of these payoff values in Nash Equilibriums are explored in detaila and best strategy which give the maximal sum of the payoffs of both players depending on the initial quntum state (strategy). The whole aspects as the function of a parameter representing itial quntum state are summarized in a figure.  We find that the structure of the relations of magnitudes among payoff functions becomes rather complicated.  If we would always choose the maximum payoff solution among them, the solution is determined uniquely, depending on  the initial quantum state, completly.  The last section 5 is devoted to discussions on some famous symmetric games with four parameters in the payoff bi-matrises and some common properties are found between them and the stag hunt game. Lastly a little remark is made on Prisoner's Dillemma.

\section{Classical stag hunt Game and its Nash Equilibrium}
In this section we make a review of a useual (classical) stag hunt game, show what is the Nash equilibrium solution in this game and explain a dilemmma of this game. This solution will be needed to compare with a quntum one of the game. In the game, both players simultaneously choose their actions (Static Game) and each player knows perfectly the values of the payoff functions of all his (her) opponents (Complete Information).

Notations in this paper is followings;\\
1. the number of players are represented by $i, \cdots, N$. In this paper $N$ is fixed to be $2$ throughout. \\
2.The set of strategies is discrete and it constitutes the strategyic space $S_i$ available to each player.\\
3. The payoff functions $ \$ _i=\$ _i (s_1,s_2, \cdots , s_N) $ assign i-th player a real number depending on the strategies chosen by his(her) opponents.  

\subsection{stag hunt Game}
The stag hunt game is a non-zero sum game and the payoff bimatrix of this game  in $2\times 2$ symmetric case is given by Fig.1, where the strategy $C$ means cooperation and the $D$ means defection\cite{rap}. Players are a man, Bob and a woman, Alice as referred in the usal game theory.  

\begin{center}
\begin{tabular}{c} 
 \hspace{5mm} Bob\\
\end{tabular}

Alice
\begin{tabular}{|l||c|c|} \hline
 &C&D\\ \hline \hline
C&(a,a)&(d,b)\\ \hline
D&(b,d)&(c,c) \\ \hline
\end{tabular}

\refstepcounter{figure}
Fig. \thefigure. Payoff function in the stag hunt game.
\addcontentsline{lof}{figure}%
{\protect\numberline{\thefigure}{Payoff function in the stag hunt game.}}
\end{center}

Here $a,b,c,d$ are real numbers and the conndition $a>b>c>d$ is required for the stag hunt game. In this paper we analytically pursue a discussion  as possible as we can and so we do not assign them any explicit values. 

This game may be interpreted as follows. 
Tow persons want to hunt a deer or a rabit. If they cooprate together in the hunting, they can get a big game, a deer and divid it into two parts. He, however,  by himself can get only a small game, a rabit. So what they cooperate is the best strategy each other. It, however, does not always follow that his opponent cooperates with him  when he will cooperate. If one person believes that his opponent always cooperates and he really cooperates with the opponent, while the opponent does not cooperate and is going to take a rabit alone, then the payoff he can get become to be the lowest value because  the opponent can get by himself a rabit but he can get nothing and so he is very chagrined. Conversely when he defects and the opponent cooperates, he can get a rabit by himself and monopolize it, but the opponent can get nothing. Then he will get the larger payoff than in the case both of them get rabit, respectively. Notice that the the payoff function of Fig.1 with the condition $a>b>c>d$ well reflects  this situation.  

\subsection{Nash Equilibrium solution}
When a payoff function is given, the game theory presents various solutions of what players should act in the game. The most significant solution in non- zero-sum games is Nash  Equilibrium. In general non-zero-sum $N$ players games, the set of strategies $(s_1^{\ast}, s_2^{\ast}, \cdots, s_N^{\ast})$ constitute Nash Equilibrium, if, for each player $i$, the strategy $s_i^{\ast}$ satisfies the following equation:
\begin{equation}
\$_i(s_1^{\ast}, s_2^{\ast}, \cdots, s_{i-1}^{\ast}, s_i^{\ast}, s_{i+1}^{\ast},\cdots, s_N^{\ast}) \geq 
\$_i(s_1^{\ast}, s_2^{\ast}, \cdots, s_{i-1}^{\ast}, s_i, s_{i+1}^{\ast},\cdots, s_N^{\ast})
\end{equation}
for every strategy $s_i$ belonging to the i-th strategic space $S_i$. So Nash Equilibrium corresonds to a set of strategies, each one representing the best choice for each single player if all his opponents take their best choice, too.

  There are two Nash Equilibriums in the stag hunt game.  They are easily found to be $(D,D)$ or $(C,C)$. 

\subsection{Mixed Classical Strategies}
In this section we study mixed classical strategies made in usal game theories. We know that when the strategies are spread to mixed strategies, there are always some Nash Equilibriums\cite{suz}. In mixed strategieas Alice takes the strategy $C$ with the possibility $p$ and $D$ with the possibility $1-p$.  Bob adopts the strategy $C$ with the possibility $q$ and $D$ with the possibility $1-q$. The conditions  $0\geq p \geq 1$ and $0\geq q \geq 1$ should be imposed on the probabilities, for they means possibility. 

Then  Nash Equilibriums in the mixed strategies can be estimated as follows.
The expectation value of Alice's reward $\$_A$ and Bob's $\$_B$ one are given by 
\begin{eqnarray}
\$_A(p,q)&=&pqa+p(1-q)d+q(1-p)b+(1-p)(1-q)c, \\
\$_B(p,q)&=&pqa+p(1-q)b+q(1-p)d+(1-p)(1-q)c. 
\end{eqnarray}
From them and the condition for Nash Equilibriums given in (1) we obtain
\begin{eqnarray}
\Delta _A \equiv \$_A(p^{\ast},q^{\ast})-\$_A(p,q^{\ast}) =
(p^{\ast} -p) \{ q^{\ast} (a-b) + (1- q^{\ast})(d-c) \} \geq 0,&& \\
\Delta _B \equiv \$_B(p^{\ast},q^{\ast})-\$_B(p^{\ast},q) =
(q^{\ast} -q) \{ p^{\ast} (a-b) + (1- p^{\ast})(d-c) \} \geq 0.&&
\end{eqnarray}
Throughout all caluclations, noticing that there is a symmetry, 
\begin{equation}
A \Longleftrightarrow B, \;\;\; b \Longleftrightarrow d, 
\end{equation}
is useful. Three solutions can be found from the equations (4) and (5).\\

[1]$p^{\ast}=q^{\ast}=1$\\
In this case two players get the same reward,
\begin{equation}
\$_A(1,1)=\$_B(1,1)=a.
\end{equation}

[2]$p^{\ast}=q^{\ast}=0$\\
In this case two players also get the same reward, 
\begin{equation}
\$_A(0,0)=\$_B(0,0)=c.
\end{equation}

[3]$p^{\ast}=q^{\ast}=\frac{c-d}{a-b+c-d}\equiv m$\\
In this case the reward of two players are
\begin{equation}
\$_A(m,m)=\$_B(m,m)=\frac{ac-bd}{a-b+c-d}>0.
\end{equation}
By calculating directly, the following relations hold good among estimated rewards;
\begin{equation}
\$(0,0)=a > b> \$(m,m) > c= \$(1,1) >d,
\end{equation}
where $\$ $ with no suffix means $\$ = \$_A = \$_B$.

\section{Quantization Scheme of Game Theory}
Firstly the quntization has been discussed by Meyer\cite{mey}. 
At the present day tow ways for the quantization of game theory has been known. One is the formulation by Eisert et al\cite{eis}. and another is the one by Marianatto et al.\cite{mar}. 
We follow the latter method since it is clear to calculate various quantities in the letter formulation and in the Battle of sexes game the situation becomes corather mplex than in the analysis made by first formulation, that is, there are infinite Nash Equilibriums\cite[du]. Our aim is to cpmpare the results discussed by Marinatto et al.\cite{mar}, where  quantum strategies can give  better results than classical strategies and discuss the universality of the superriority of quantum strategies to  classical ones.

many studies by using the method has not been made yet. In the next subsection we present the formulation given by Marianotto et al.\cite{mar} in the case oh the stag hunt game.

The quntization of game theory proposed by Marinotto et al. \cite{mar}. consists of the following steps.\\

[1]Unlike in the classical game theory, we first fix an arbitrary initial quantum state belonging to the Hilbert space $S=S_A \otimes S_B$, obtained as a direct product of the two strategic spaces of the two players, whose orthognal bases consist of the vectors associated to the pure strategies\footnote{Notice that Marinotto et al. used the word 'strategy' in an unusal meaning. They call the initial state in a game 'strategy'.} classically.\\

[2] Each player can manipulate the initial state vector (strategy) by performing some local transformation to obtain a suitable final state vector which will represent the quntum strategy the two players are going to play\footnote{ In the language employed by Marinotto et al.  the word 'strategy' coresponds to the initial state and the word 'tactics' means the operators related to the transformation.}\\

[3]The expected payoff of each player have to be evaluated by calculating the squared modula of the projections of the final quantum state onto the basis vectors of space $S$, and then by adding up the obtained values multiplied by the appropriate payoff coefficients deducible from the payoff bimatrix. \\

[4]Each player has to eventually play the clasicall pure strategy, which results a measurement process on the final strategy, that is, a projection onto the basis vectors. 

These process are summarized in Fig.2.

In calculating explicitly the physical quanties I follow density matrix approach introduced by Marinotto et al.\cite{mar}. In the case of the stag hunt game the basis vectors are 
\begin{equation}
|C>,\;\;\;\;|D>,
\end{equation}
where $|C>$ represent a cooperate state and on the other hand $|D>$ represent a defect state, respectively.
Let introduce a unitary and hermitian operator $U$ interchange vectors  $|C>$ and $|D>$\footnote{ we can really show that such a situation is possible by explicitly giving  the representation of the state vectors and the operator: for example it is done by choose 
$|C>=\left[ \begin{array}{c}
1\\
0
\end{array}\right]$, 
$D=\left[ \begin{array}{c}
0\\
1
\end{array}\right]$ 
and $U=\sigma _x$, the first Pauli matrix. };

\begin{eqnarray}
U|C>&=&|D>, \nonumber \\
U|D>&=&|C>, \nonumber \\
C^{\dagger}=C&=&C^{-1}.
\end{eqnarray} 
The our assumption is that each palyer can modify his own strategy by applying to his reduced part of the total density matrix $\rho _{in}$, which represents the initial state of the game, so transformed (final) density matrix is given by\begin{eqnarray}
\rho_{fin}^{A}&=&[p I \rho_{in}^{A} I^{\dagger} +(1-p)U \rho_{in}^{A}U^{\dagger}],\\
\rho_{fin}^{B} &=&[q I \rho_{in}^{B} I^{\dagger} +(1-q)U \rho_{in}^{B}U^{\dagger}
\end{eqnarray}
where Alice acts with the identity $I$ with probability $p$ and with $U$ with probability $(1-p)$ and Bob acts the same operator with probabolity $q$ and $(1-q)$, respectively. 
From equation (13) and (14) the following expression for the final density matrix is obtained;
\begin{eqnarray}
\rho_{fin}& =&pq (I_A \otimes I_B)\rho_{in}  (I_A^{\dagger} \otimes I_B^{\dagger}) +p(1-q) (I_A \otimes U_B)\rho_{in}  (I_A^{\dagger} \otimes U_B^{\dagger}) \nonumber \\
&&+  (1-p)q (U_A \otimes I_B)\rho_{in}  (U_A^{\dagger} \otimes I_B^{\dagger})
\nonumber \\
&&+(1-p)(1-q) (U_A \otimes U_B)\rho_{in}  (U_A^{\dagger} \otimes U_B^{\dagger}).\end{eqnarray}

In order to calculate the payoff functions, we introduce the following two operators;
\begin{eqnarray}
P_{A}=a|CC><CC| + c(|CD><CD|+|DC><DC|)+b|DD><DD|,&&\\
P_{B}=b|CC><CC| + c(|CD><CD|+|DC><DC|)+a|DD><DD|.&&
\end{eqnarray}
After all, the payoff functions can be evaluated by employing these operators;
\begin{eqnarray}
\$_{A}& =&Tr (P_A \rho_{in}),\\
\$_{B}& =&Tr (P_B \rho_{in}).
\end{eqnarray}

Why this formulae for payoff functions corresponds to the above mentioned quantization formulation is explicitly given by Marinatto et al.\cite{mar}. 
Eventually in this formalisum,  th tactics or local operator corresponds to the special unitary matrix $U\in SU(2)$;
\begin{equation}
 U=\left[ \begin{array}{cc}
\sqrt{p} & ie^{i\phi} \sqrt{1-p}\\
ie^{i\phi} \sqrt{1-p}&\sqrt{p}
\end{array}\right],
\end{equation}
where $\phi$ is an arbitrary phase factor\footnote{This form do not cover all the ranges of $SU(2)$ and that general $SU(2)$ matorix should be considered as tacticses is pointed out\cite{du, ben}}.

\section{Quantization of stag hunt Game and its Nash Equilibrium}
I explicitly calculate the payof functions in the case of the deer huntion game according to the formulation of the previous section. Moreover we show that there are three Nash Equilibriums in the game and make a discuss on the properties of them in datails.  

\subsection{The case of Non-entanglrd Strategy}
In this section we discuss the case where an initial state (strategy) is a factorizable state. When started with the factrizable state, in all game studied by using both formulations of quantization of game theory, it is known that the results are the same as in the case of the classical analysis.

We start with an initial state $|CC>$\footnote{it is not significant which factrizable state is taken as an initial strategy since all factorizable states can be given by applying an appropriate unitary transformation on an arbitraly factorizable state.}.  Then we obtain from equation (15);
\begin{eqnarray}
\rho_{fin}&=&pq|CC><CC| + p(1-q)|CD><CD|+(1-p)q|DC><DC|\nonumber \\
&&+(1-p)(1-q)|DD><DD|.
\end{eqnarray}
We can find three Nash Equilibriums same as ones in the classical case.\\

[1] $p=q=1$\\
In this case $\rho _{fin} $ is given by 
\begin{equation}
\rho_{fin}=|CC><CC| 
\end{equation}
and the payoff is evaluated from equations (18) and (19);
\begin{equation}
\$_{A}= \$_{B}=a.
\end{equation}

[2]$p=q=0$\\
In this case $\rho _{fin} $ is given by 
\begin{equation}
\rho_{fin}=|DD><DD| 
\end{equation}
and the payoff is evaluated from equations (18) and (19);
\begin{equation}
\$_{A}= \$_{B}=c.
\end{equation}

[3]$p=q=m$\\
$m$ is the same value as the one given in [3] in the section 2.3. 
 In this case $\rho _{fin} $ is given by 
\begin{eqnarray}
\rho_{fin}&=&m^2|CC><CC| + m(1-m)(|CD><CD|+|DC><DC|)\nonumber \\
&&+(1-m)(1-m)|DD><DD|
\end{eqnarray}
and the payoff is given by  
\begin{equation}
\$_{A}= \$_{B}=\frac{ac-bd}{a-b+c-d}>0.
\end{equation}

As we expect all the results are the same as the ones in the classical case.

\subsection{The case of entangled Strategy}
Useually the most interesting case is when  entangled states, which is essentialy the peculiar to quntum mecanics\cite{sak},  are taken as an initial state.
We begin with general entangled state 
\begin{equation}
|\psi _{in}>= \alpha |CC>+ \beta |DD>\;\;\; with\; |\alpha|+|\beta|=1.
\end{equation}
Then by calculating directly from definitions, we get 
\begin{eqnarray}
\rho_{in}&=&|\alpha|^2|CC><CC| + \alpha \beta^{\ast}|CC><DD| \nonumber \\
    &&+\alpha ^{\ast}\beta |DD><CC|)+|\beta |^2|DD><DD|,\\
P_A&=&a|CC><CC| + d|CD><CD|+b|DC><CC|) \nonumber \\
&&+c|DD><DD|,
\end{eqnarray}
and
\begin{eqnarray}
\rho_{fin}&=&(pq+|\beta |^2(1-p-q))|CC><CC|+ (pq+|\alpha |^2(1-p-q))|DD><DD|
\nonumber \\
&&+(\alpha \beta ^{\ast}pq+|\alpha |^{\ast}b(1-p)(1-q))|CC><DD|
\nonumber \\
&&+(\alpha \beta ^{\ast}(1-p)(1-q)+|\alpha |^{\ast}bpq)|DD><CC|
\nonumber \\
&&+(-pq+|\alpha |^2q+ |\beta |^2 p)|DC><DC|
\nonumber \\
&&+(-pq+|\alpha |^2p+ |\beta |^2 q)|CD><C|
\nonumber \\
&&+(\alpha \beta ^{\ast}p(1-q)+|\alpha |^{\ast}b(1-p)q)|CD><DC|
\nonumber \\
&&+(\alpha \beta ^{\ast}q(1-p)+|\alpha |^{\ast}b(1-q)p)|DC><CD|.
\nonumber \\
\end{eqnarray}
By using  equations (18) and (19), we can evaluate the payoff functions;
\begin{eqnarray}
\$_A&=&pq(a+c-b-d)+p(|\alpha|^2(d-c)+|\beta|^2(b-a))\nonumber\\
&&+q(|\alpha|^2(b-c)+|\beta|^2(d-a))+a|\beta|^2+c |\alpha |^2,\\
\$_B&=&pq(a+c-b-d)+p(|\alpha|^2(b-c)+|\beta|^2(d-a))\nonumber\\
&&+q(|\alpha|^2(d-c)+|\beta|^2(b-a))+a|\beta|^2+c |\alpha |^2.
\end{eqnarray}
So we obtain
\begin{eqnarray}
\Delta _A &\equiv &\$_A(p^{\ast},q^{\ast})-\$_A(p,q^{\ast})\nonumber\\
&=&(p^{\ast}-p) \left[ q^{\ast} (a+c-b-d)+ |\alpha|^2(d-c)+|\beta|^2(b-a) \right],\\
\Delta _B &\equiv &\$_B(p^{\ast},q^{\ast})-\$_B(p^{\ast},q)\nonumber\\
&=&(q^{\ast}-q) \left[ p^{\ast} (a+c-b-d)+ |\alpha|^2(d-c)+|\beta|^2(b-a) \right].
\end{eqnarray}
From these, we can find three Nash Equilibriums.\\

[1]$p^{\ast}=q^{\ast}=1$\\
The payoff function is given by
\begin{equation}
\$_A (1,1)=\$_B(1,1)= a|\alpha |^2+ c |\beta |^2.
\end{equation}

[2]$p^{\ast}=q^{\ast}=0$\\
The payoff function is given by
\begin{equation}
\$_A (0,0)=\$_B(0,0)= c|\alpha |^2+ a |\beta |^2.
\end{equation}

[3]$p^{\ast}=q^{\ast}=m_q$\\
The definition of $m_q$ is 
\begin{equation}
m_q=\frac{(c-d)|\alpha |^2+ (a-b) |\beta |^2}{a-b+c-d}.
\end{equation}
After long and tedious calculation, the payoff function is given by
\begin{equation}
\$_A (m_q, m_q)=\$_B(m_q,m_q)= \frac{(ac-bd)+|\alpha |^2 |\beta |^2 (a+b-c-d)(a-b-c+d)}{a-b+c-d}.
\end{equation}

When we take $\alpha =1$ or $\beta=1$, it can be seen easily that the peculiar property different from the Battle of the Sexes game appears. 
Then we find the following relation;
\begin{eqnarray}
\$ (1,1)=c< \$(m_q, m_q) =\frac{ac-bd}{a-b+c-d}<\$(0,0)=a,\;\;\;for |a|=0,\\
\$ (0,0)=c< \$(m_q, m_q) =\frac{ac-bd}{a-b+c-d}<\$_B(0,0)=a,\;\;\;for |b|=0.
\end{eqnarray}
Such a relation does not appear in the Battle of the Sexes game in any parameter regions at all.  
On the other hand we can get the same relation as the case of the Battle of the Sexes game when we choose $|\alpha |= |\beta |= 1/\sqrt{2}$;
\begin{equation}
\$ (0,0)=\$(1,1)= \frac{a+c}{2} > \$ (m_q, m_q) = \frac{a+b+c+d}{2}.
\end{equation}
In the battle of the Sexes game, this relation always hold good in all the parameter gegion\cite{mar}.

From these situation, we wonder whether there is the case such as $\$ (0,0),\;\$(1,1)< \$ (m_q, m_q) $ and furthermore we sholuld generally investigate  the all relations standing for $\$ (0,0),\;\$(1,1),\;\$ (m_q, m_q)$ in the all parameter regions for $| \alpha, \; \beta,\; a,b,c,d$ systematically. For the purpose, we calculate explicitelly $\Delta _{(1,1)}\equiv \$(1,1)- \$(m_q, m_q)$ and  $\Delta _{(0,0)}\equiv \$(0,0)- \$(m_q, m_q)$. \\

After long but straightforward calculation, we obtain the following form for $\Delta (1,1)$;
\begin{equation}
\Delta _{(1,1)}= C_1X^2+B_1X+A_1,
\end{equation}
where 
\begin{eqnarray}
C_1&=&\frac{(a+b-c-d)(a-b-c+d)}{a-b+c-d},\\
B_1&=&a-b+C_1,\\
A_1&=&\frac{(b-c)(c-d)}{-a+b-c+d}<0,\\
X&=&|\alpha |^2.
\end{eqnarray}

The solutions  of equation $\Delta _{(1,1)}= C_1X^2+B_1X+A_1=0$ turn to be 
\begin{eqnarray}
x_+^{(1)}&=&\frac{(b-c)}{a+b-c-d},\\
x_-^{(1)}&=&\frac{c-d}{-a+b+c-d}.
\end{eqnarray}
Moreover noticing that 
\begin{eqnarray}
0<x_+^{(1)}<1,&&0\leq X \leq 1\\
\Delta _{(1,1)}(X=0)&=&A_1 <0,\\
\Delta _{(1,1)}(X=1)&=&A_1+B_1+C_1 >0,
\end{eqnarray}
and using the standard results of the theory of quadratic equation, we can obtain the following relations for sign of $\Delta _{(1,1)}$;
\begin{eqnarray}
X> x_+^{(1)} &\Longrightarrow& \Delta _{(1,1)}>0,\nonumber \\
X= x_+^{(1)} &\Longrightarrow& \Delta _{(1,1)}=0,\nonumber \\
X< x_+ ^{(1)}&\Longrightarrow& \Delta _{(1,1)}<0.
\end{eqnarray}

For $\Delta (0,0)$, we can lead to the following equation;
\begin{equation}
\Delta _{(0,0)}= C_0X^2+B_0X+A,
\end{equation}
where 
\begin{eqnarray}
C_0&=&\frac{(a+b-c-d)(a-b-c+d)}{a-b+c-d}=C_1,\\
B_0&=&a-b+C_1=B_1,\\
A_0&=&\frac{(a-b)(a-d)}{a-b+c-d}>0.
\end{eqnarray}

The solutions  of equation $\Delta _{(0,0)}=0$ turn to be 
\begin{eqnarray}
x_-^{(0)}&=&\frac{(a-d)}{a+b-c-d},\\
x_+^{(0)}&=&\frac{a-b}{a-b-c+d}.
\end{eqnarray}
Moreover noticing that 
\begin{eqnarray}
0<x_-^{(0)}<1,&&0\leq X \leq 1\\
\Delta _{(0,0)}(X=0)&=&A_0 >0,\\
\Delta _{(0,0)}(X=1)&=&A_0+B_0+C_0 <0,
\end{eqnarray}
and by making paralell argument with the case of  $\Delta _{(1,1)}$, we can obtain the relations for $\Delta _{(0,0)}$;
\begin{eqnarray}
X> x_-^{(0)} &\Longrightarrow& \Delta _{(0,0)}<0,\nonumber \\
X= x_-^{(0)} &\Longrightarrow& \Delta _{(0,0)}=0,\nonumber \\
X< x_-^{(0)} &\Longrightarrow& \Delta _{(0,0)}>0.
\end{eqnarray}

Consequently the relative magnitudes of payoff functions as $|\alpha |^2$ become larger are summerized as follows.
\begin{eqnarray}
(1)\$ (0,0) > \$ (m_q,m_q) > \S(1,1)&& for \;\; 0\leq |\alpha |^2 < x_+^{(1)}, \nonumber \\
(2)\$ (0,0) > \$ (m_q,m_q) = \S(1,1)&& for \;\; |\alpha |^2 =x_+^{(1)}, \nonumber \\
(3)\$ (0,0) > \S(1,1) >\$ (m_q,m_q) && for \;\; x_+^{(1)}<|\alpha |^2 <\frac{1}{2} , \nonumber \\
(4)\$ (0,0) =\S(1,1) >\$ (m_q,m_q) && for \;\; |\alpha |^2 =\frac{1}{2}, \nonumber \\
(5)\$ (1,1) > \S(0,0) >\$ (m_q,m_q) && for \;\; \frac{1}{2}<|\alpha |^2 <x_-^{(0)} , \nonumber \\
(6)\$ (1,1) > \S(0,0) =\$ (m_q,m_q) && for \;\;|\alpha |^2 =x_-^{(0)}, \nonumber \\
(7)\$ (1,1)  >\$ (m_q,m_q) > \S(0,0)&& for \;\; x_-^{(0)}<|\alpha |^2 \leq 1. \nonumber 
\end{eqnarray}
This results are consistent with the preliminary arguments in equations () and (). Marinatto et al.\cite{mar} conclude that we there is a unique Nash Equilibrium at $|\alpha |^2= | \beta | ^2= \frac{1}{2}$ and the Dilemma in this game are resolved. It, however, has been pointed out that the degeneracy of Nash Equilibrium is not resolve completly. The communications between two players are needed to resolve the degeneracy\cite{ben}.

Since four parameters are needed in the stag hunt game not but three in the Battle of the Sexes game, the structure of the relations representing the relative magnitudes among payoff functions becomes rather complex. While $\$ _a\neq \$ $in the Battle of the Sexes game, $\$ _A= \$ _B$ satisfies for any $p, \; q$ in the stag hunt game and the situation in stag hunt game is essentially different from the one in the Battle of the Sexes game. In the stag hunt game the same strategies are always taken but three solutions exist for any $p,\; q$. If we would always choose the maximum payoff solution among them, the solution is determined uniquely. Aspecially at  $|\alpha |^2= | \beta | ^2= \frac{1}{2}$, the same situation as the Battle of the Sexes game occurs in the stag hunt game. At the point we really see $ \$ (0,0)=\$ (1,1) > \$ (m_q, m_q)$ like the Battle of the Sexes game. Generally, however,  the strategy with the best payoff depends on the initial quantum state, that is,  $|\alpha |,\; | \beta | $. There are not the case the mixed strategy give best payoff as in the classical case.
The classical case corresponds to the case with $|\alpha |^2= \frac{1}{2}$ in the quntum case where $p=q=1$ give the best payoff for both players. On the other hand at $|\alpha |^2<  \frac{1}{2}$bin quantum case, the best payoff comes when  $p=q=0$ and such situation never occur classically!. Which strategy can give best payoff depends on the given initial state, completly.  

\addtocounter{figure}{-1}
\begin{figure}[htbp]
\begin{center}
\includegraphics{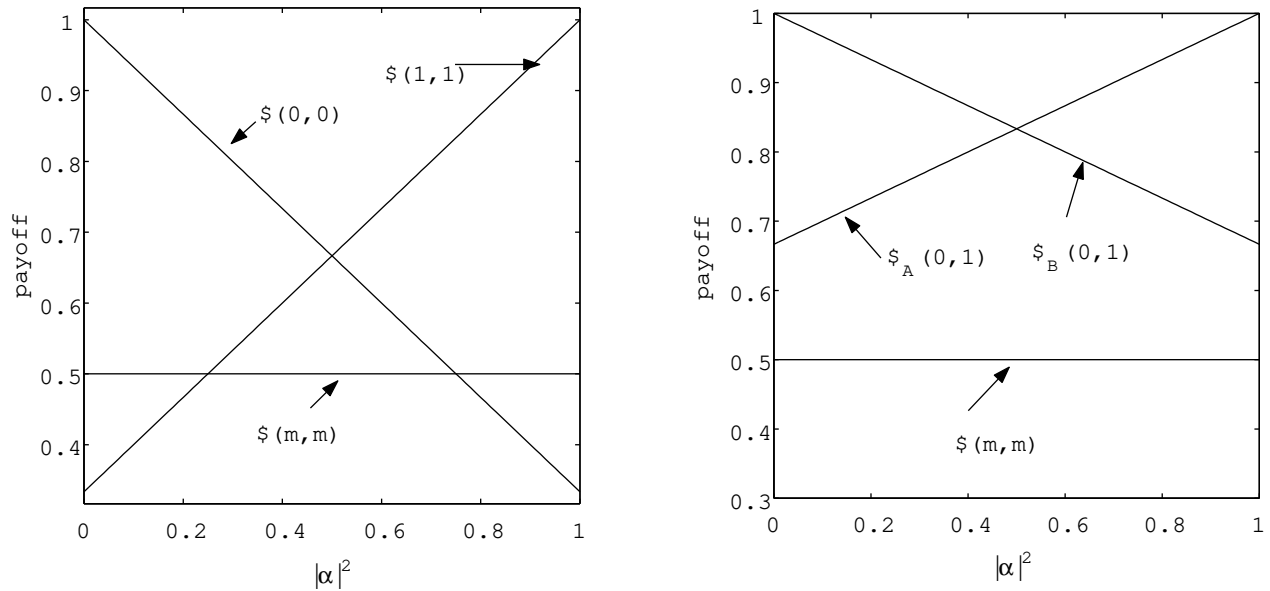}
\end{center}
\caption{Payoff of the stag hunt game vs. $|\alpha |^2$ (left side graph) and payoff of the Leader game vs. $|\alpha |^2$ (right side graph).}
\label{fig:pay3.esp-2}
\end{figure}

\section{Discussion}
By quntizing the stag hunt game along Marinott\cite{}, we find that the pecurial strategy give the best payoff which the situation never occurs in classical treatment and clear the structure of Nach Equilibrium solutions as function the initial quantum state $|\alpha |, \; | \beta | ^2= \frac{1}{2}$. The structure is rather complex since the stag hunt game has four parameters in the payoff bi-matrix while  there are three parameters in taht of the Battle of the Sexes game.\\
We give the general forms of payoff function of both player in $2\times 2$  symmetric bi-matrix game with the four parameters where is the bonus in this paper and so we can discuss general games with this form. Amon them the intersting games referred in many books of game theory are 
\begin{eqnarray} 
(i)&&Chicken \;game; \;\;\; b>a>d>c,\\ 
(ii)&&Leader \; game; \;\;\; b>d>a>c,\\ 
(iii)&&Secrete \;meeting \;game; \;\;\; d>b>a>c,\\ 
(iv)&&Prisoner's \;Dilemma \;game; \;\;\;  b>a>c>d,
\end{eqnarray} 
where the inequalitys show the order of the magnitude in each payoff bi-matrix. We briefly discuss the four games with the special case that the biggest value in the payoff  bi-matrix is normalized to be  $1$ and the smallest one is to be $0$. Furthermore The second bigger value is taken $\frac{2}{3}$ and the third bigger value is taken $\frac{1}{3}$. The values of the payoff bi-matrix are arranged  at regular intervals between $0$ and $1$.

First three games (Chickin, Leader and Secrete meeting games) have three Nach Equilibriums in the same way as their classical versions. They are (i)$p=1,\;q=0$, (ii)$p=0,\;q=1$ and (iii)$p=q=\frac{d-c}{b+d-a-c}$. The Prisoner's Dilemma game
shows very different aspect from these three games and will be discussed separately from them. When the special values of the payoff bi-matricses, such as introduced in the above discussion,  payoff function v.s. $|a|^2$ are given in table 1 and their values  v.s. $|a|^2$ are given in the left hand side of the figure 1 (Chickin, and Secrete meating games) and the right hand side in the figure 1 (Leader game) where each sloping line must be exchanged in $(p,q)=(1,0)$. The same notation should be applied for the left hand side of figure 1 in Chickin and Secrete meating games. The essential results are same as the one of the stag hunt game and the difference between the figure 2 and 3 depens on how the parameters in payoff functions are chosen and is not crusial.   
\begin{center}
\begin{tabular}{|l|c|c|c|}\hline
&\makebox[30mm]{$p=1,\;q=0$}&\makebox[30mm]{$p=1,\;q=0$}&\makebox[30mm]{$p=q=m$}
\\ \hline 
 Chickin Game&($\frac{|a|^2}{3}+|b|^2,|a|^2+\frac{|b|^2}{3}$)&($|a|^2+\frac{|b|^2}{3},\frac{|a|^2}{3}+|b|^2$)&($\frac{1}{2},\frac{1}{2})$\\ \hline
 Leader Game&($\frac{2|a|^2}{3}+|b|^2,|a|^2+\frac{|b|^2}{3}$)&($|a|^2+\frac{2|b|^2}{3},\frac{2|a|^2}{3}+|b|^2$)&($\frac{1}{2},\frac{1}{2})$\\ \hline
 Secrete Meeting Game&($|a|^2+\frac{2|b|^2}{3},\frac{2|a|^2}{3}+|b|^2$)&($\frac{2|a|^2}{3}+|b|^2,|a|^2+\frac{2|b|^2}{3}$)&($\frac{1}{2},\frac{1}{2})$\\ \hline 
\end{tabular}
\\
Table 1. Payoff functions of Nash Equilibrium in the symmetric games with four parameters in bi-matrix where the parameters are taken to be $1,2/3,1/3,0$ in orderof the magnitude. The first values in the parentheses mean the payoff values of player A and second one mean the ones of player B.\\
\end{center}

At $|a|^2=|b|^2=\frac{1}{2}$, which is maximal entangled sate, in the all three games, Nash Equilibrium point corresponding to a mixed strategy is equally to be $m=\frac{1}{2}$ same as the classical case! At this point, $\$(1,0)=\$(0,1)\geq \$(m,m)$ and  Nash Equilibrium $p=1,\; q=0$ and  $p=0,\; q=1$ are better than another Nash Equilibrium $p=q=m$ in the view of payoff. When getting out of this point means, the degeneracy of $p=1,\; q=0$ and  $p=0,\; q=1$ is resolved.   

Totally these  games with four parameters in the payoff bi-matrix including tha stag hunt game have three Nash Equilibriums and the best strategy among them in the sense that the sum of the payoffs of both players become maximum strongly depends on the quntum initial state (strategy). This aspect is drastically different from clasical situations. 

Lastly  we make a smallremark on  Prisoner's Dilemma whic has a dominated strategy (defect,Defect).  Which strategy is best depends on the  quntum initial state (strategy) intricately even when the parameters are taken to be fix to be a special value stisfying $b+d<2a$ and a dominanated strategy is not always best strategy. For example in the case of $b=1, a=5/6, c=1/3, d=0$, a mixed strategy $(p, q) neq (0 ,1) \; and \; (1,0) $ becomes best strategy in the view of paypff function at $| \alpha |\sim 0$ or $| \alpha | \sim 1$. The details seems to depend strongly on the various parameter and the analysis wmay be very intricate. So the discussion on the complete analysis for  Prisoner's Dilemma is beyond this paperwhose aim mainly is to discuss stag hunt game.


\end{document}